# CUTTING THROUGH THE NOISE: AN EMPIRICAL COMPARISON OF PSYCHO-ACOUSTIC AND ENVELOPE-BASED FEATURES FOR MACHINERY FAULT DETECTION

*Peter Wißbrock[1], Yvonne Richter[2], David Pelkmann[2], Zhao Ren[3], Gregory Palmer[3]*

[1]Lenze SE, Innovation, Aerzen, Germany
peter.wissbrock@lenze.com
[2]Institut für Systemdynamik und Mechatronik, Bielefeld University of Applied Sciences, Germany
[3]L3S Research Center, Leibniz University of Hannover, Germany

## ABSTRACT

Acoustic-based fault detection has been one of the key instruments to monitor the health condition of mechanical parts. However, the background noise of an industrial environment may negatively influence the performance of fault detection. Limited attention has been paid to improving the robustness of fault detection against industrial environmental noise. Therefore, we present the Lenze production background-noise (LPBN) real-world dataset and an automated and noise-robust auditory inspection (ARAI) system for the end-of-line inspection of geared motors. An acoustic array is used to acquire data from motors with a minor fault, major fault, or which are healthy. A benchmark is provided to compare the psychoacoustic features with different types of envelope features based on expert knowledge of the gearbox. To the best of our knowledge, we are the first to apply time-varying psychoacoustic features for fault detection. We train a state-of-the-art one-class-classifier, on samples from healthy motors and separate the faulty ones for fault detection using a threshold. The best-performing approaches achieve an area under curve of 0.87 (logarithm envelope), 0.86 (time-varying psychoacoustics), and 0.91 (combination of both).

## 1. INTRODUCTION

Fault detection of mechanical systems like gears and motors is an important topic for industrial production. Vibration signals have been widely used to analyze the health condition of mechanical systems [1] e.g., to analyze the product quality at the end of assembly line (EoL) [2, 3]. However, for custom-built geared motors auditory EoL-inspection is still often carried out by humans, which results in inconsistent product quality assessments [4]. In previous work [5], we conceptualized an automated process for the EoL-inspection in a highly variant production of custom-built geared motors. This cannot be done economically using vibration sensors due to the amount of time required for mounting the sensors, inconsistent sensor placement, fault-depended positioning, and limited transfer frequencies [6]. An alternative is to utilize acoustic signals. However, acoustic-based fault detection has received limited attention, due to the risk of background noise in an industrial environment causing misclassification [3]. Nevertheless, acoustic signals have many advantages compared to vibration signals, while capable of delivering a comparable performance [1, 7]. For instance, no connection is needed between the motor and the sensor, and acoustic signals cover more types of faults including aerodynamic components [8]. In previous work [5] we discussed the problem of small imbalanced datasets due to the fact that samples from faulty motors are very rare. Therefore, we focus on a fault detection approach based on a one-class classifier (OCC), trained on samples of healthy motors only.

In this paper, we show that samples of motors with faults can be separated from samples of healthy motors in a noisy real-world EoL-inspection scenario. To the best of our knowledge, we are the first to evaluate acoustic-based fault detection of custom-built geared motors under industrial noise. Our contributions can be summarized as follows:

1) We present the Lenze production background-noise (LPBN)[1] real-world acoustic dataset. LPBN contains 138 samples of custom-built geared motors divided into motors labeled with *major fault* (does not fulfill standard quality), *minor fault* (fulfills standard quality), and *healthy*. This includes 43 samples in which typical industrial background noises were deliberately generated.

2) We present an automated and noise-robust auditory inspection (ARAI) system. To overcome the problem of background noise, we use a 64-microphone acoustic array, which reduces noise in the high-frequency area, and a digital filter to suppress low frequencies. This preprocessing supports our fault detection, consisting of noise-robust features and a state-of-the-art OCC, the bagging random minor.

3) We present a benchmark for gear fault detection including different definitions of the well-established envelope spectrum [1]. For LPBN, the logarithm envelope spectrum (LES) outperforms the other definitions. Additionally, our experiments show that time-varying psychoacoustic (TVPA) features outperform stationary psychoacoustic (SPA) ones. We combine features based on TVPA and LES, which achieve the best performance, with an area under curve of the receiver operating characteristic for the detection of minor faults of 0.85 and major faults of 0.91.

## 2. RELATED WORK

**Envelope Spectrum.** One of the most widely used approaches for the detection of gear and bearing faults is to use envelope-based features extracted from vibration or acoustic signals [1]. [9] showed, that the squared envelope spectrum (SES) outperforms the normal envelope spectrum (NES) with respect to the background noise.

---
Acknowledgment: Parts of this article were supported by the Ministry of Economic Affairs, Innovation, Digitization, and Energy of North Rhine Westphalia through the excellence cluster itsOWL in the projects "PsyMe" and "ML4Pro[2]" and the German Federal Ministry for Economics and Climate Action (BMWK) through the research project "IIP-Ecosphere", via funding code 01MK20006A.
[1]Part of the data will be provided by request (for research only)

Contrastively, [10] discussed vibration-based bearing fault detection under impulsive noise induced by an ore crusher. It was shown that the LES is more robust compared to the SES. To extract features from the envelope spectrum, expert knowledge of gearbox construction is necessary. In contrast, condition indicators like statistical features and time-synchronous averaging (TSA) can be used without expert knowledge [11]. However, TSA cannot be used for most motors as no position-feedback system is installed.

**Psychoacoustics.** Psychoacoustics describes human hearing and can be computed for stationary (distribution does not depend on time) or time-varying signals differently [12]. [4, 13] used a laboratory acoustic dataset of gear faults and showed that SPA outperforms statistical indicators for EoL-inspection. However, the authors suggested the use of a noise-reducing test chamber in the production. To circumvent the problem of background noise, [3] applied SPA to a vibration signal, which performed equally to condition indicators based on TSA.

**Limitations.** Both, using vibration sensors or a test chamber would lead to significantly higher costs. Therefore, we propose an auditory fault detection approach that is robust to industrial background noise. In contrast to related work [3, 4, 13], the envelope approach is considered for EoL-inspection, given that a manufacturer of gearboxes can provide constructional details of the gear. Further, the use of TVPA is not discussed in related work. A discussion of the robustness of psychoacoustics and the envelope approach is required. As no public acoustic dataset is available, we gather a representative group of samples as a starting point.

## 3. THEORETICAL BACKGROUND

**Envelope Approach.** The widely used variations normal and squared envelope spectrum [1] as well as the logarithm envelope spectrum [10] are defined by formulas (1-3), respectively. $\mathcal{F}$ is the Fourier transform, $\mathcal{H}$ is the Hilbert transform and $X[n]_{L,H}$ is the band-pass filtered time signal with lower and higher cutoff frequencies $L$ and $H$.

$$NES[f] = \left|\mathcal{F}\{|\mathcal{H}\{X[n]_{L,H}\}|\}\right| \quad (1)$$

$$SES[f] = \left|\mathcal{F}\left\{|\mathcal{H}\{X[n]_{L,H}\}|^2\right\}\right|^2 \quad (2)$$

$$LES[f] = \left|\mathcal{F}\left\{\log\left(|\mathcal{H}\{X[n]_{L,H}\}|^2\right)\right\}\right|^2 \quad (3)$$

A pitting fault at rotating mechanical parts, such as gears and bearings, leads to impulses that stimulate structural resonances [6]. The repetition frequencies of impulses are referred to as fault frequencies (FF) in the spectrum and depend on the gear construction. From the shaft frequencies $f_n$ which is the speed of the motor and gear shafts, the $k$ harmonics $FF_{shaft,n,k}$ are extracted. The $z$-harmonic of $f_n$ is the gear mesh frequency $FF_{gear,s}$ and $z \pm 1$-harmonics are the sidebands $FF_{side,\pm n}$, where $z$ is the number of teeth and $s$ is the gear stage. All indices are numbered from the motor shaft to the output shaft.

$$FF_{shaft,n,k} = k \cdot f_n \; for \; k \in \{1,2,3,4,\dots\} \quad (4)$$
$$FF_{gear,s} = f_n \cdot z_{2s-1} \; for \; n = s \in \{1,2,\dots\} \quad (5)$$
$$FF_{side,\pm n} = FF_{gear,s} \pm f_n \; \forall \; n \in \{2s, 2s-1\} \quad (6)$$

**Psychoacoustics.** Methods were developed to describe how sound is perceived by humans and to objectively quantify sound [12], including loudness, roughness, and fluctuation. Loudness describes the perceived sound level. Loudness perception is a function of sound pressure level, frequency, and spectral shape of the sound [13]. Stationary loudness covers those dependencies over the entire time signal. As per norm [14], time-varying loudness considers that the loudness is different for each timeframe of 2 ms. Roughness describes the subjective perception of fast amplitude modulation and thus characterizes the dynamic noise over the temporal deviation of the loudness spectrum with modulation frequencies from 20 to 300 Hz. As defined in [15] time-varying roughness is computed for timeframes of 200 ms. Similar to the roughness, the fluctuation describes the loudness modulation, but focusing on modulation frequencies lower than 20 Hz and its time-varying representation has the same time resolution as loudness [12].

## 4. AUTOMATED ROBUST AUDITORY INSPECTION

ARAI, shown in Fig. 1, consists of an acoustic array, which suppresses background noise during the measuring. However, the directional efficiency is lower for low frequencies. Therefore, a band-pass filter suppresses both, low frequencies and the area of the switching frequency induced by the frequency inverter. A filter with finite impulse response (FIR) is used twice, once for each forward and backward pass respectively, with a window length covering 7.5 periods of lower cutoff frequency $L$. Various signal transformation and feature extraction approaches can be applied to ARAI. Fig. 1 shows the approaches which are applied to LPBN in the next section. For the envelope spectrum formulas (1-3) are used. Then the expert features are defined as the maximum amplitude in the envelope spectrum at the position of FF. To circumvent the problems of inaccurate motor speed and discrete frequency position, a tolerance of $\pm 1\%$ is applied to FFs positions. For psychoacoustics, no expert features are necessary, and SPA can be used directly as features. However, as fault impulses are of short nature (refer to Fig. 5), TVPA is discussed comparatively. Statistical features are extracted from TVPA, which is described in the next section. To compare different feature vectors for fault detection, a one-class classification approach is used. An OCC is trained on data from one class, here the healthy labeled one. In the classification phase, it outputs a similarity to that class, which is observed over a threshold to determine whether the presented data belongs to the trained class.

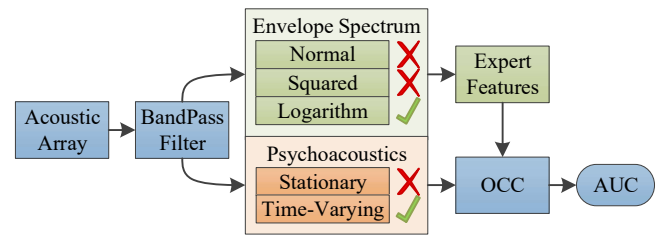

**Fig. 1:** An overview of ARAI and the compared signal transformation approaches. Steps for the envelope spectrum are in green, psychoacoustics in orange, and general steps in blue. Crosses mark features that are not noise-robust for the LPBN dataset.

## 5. LENZE PRODUCTION BACKGROUND-NOISE

**Dataset Construction.** Fig. 2 shows the measuring system, which was installed as a prototype in a production line for geared motors. The motors are operated for five seconds with rated speed. Once the rated speed is reached, an acoustic array of type SoundCam 2.0 [16] is used to measure one channel of directed sound. It consists of 64 microphones and has a half-power beamwidth of 800 Hz.

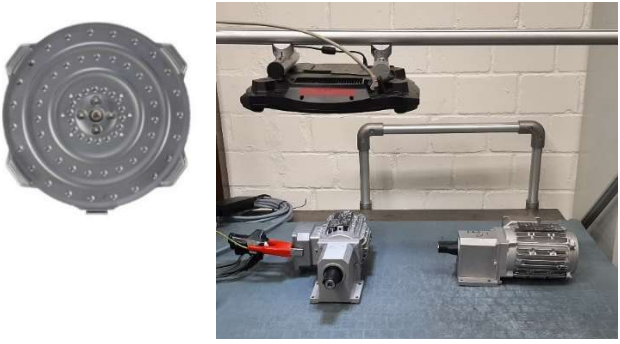

**Fig. 2:** EoL-measuring system, used to collect LPBN dataset with acoustic array (left and top) and device under test (bottom).

During the data acquisition period, samples are taken from 78 3~AC-motors with a power of 90 W, 1375 rpm of rated speed, and helical gearbox with a ratio of 10 using 2 gear stages $s$. Those motors are labeled by 3 reviewers from industry and academia with a background in EoL-inspection, psychoacoustics, and data analytics. The sound quality is inspected to determine if the motors are healthy, have minor, or major faults. To compare these labels, Kendall's coefficient of concordance $W$ [17] is computed to 0.85, while the best would be 1 and worse 0. The labels of the three reviews are then aggregated by majority voting. Most of the labeled faults are identified as gear pitting by experts, but the other faults are also circumferential and related to the considered fault frequencies.

To present a real-world scenario, the OCC is trained using a set of 56 samples that fulfill standard quality. This training dataset includes 25 motors with minor faults and 31 healthy ones. The samples are contaminated by acoustic noise observed in the production hall. To evaluate the robustness, 82 samples are used as the test dataset including 22 samples of motors with major faults. Additionally, in total 60 samples are taken with four healthy motors, two with minor faults, and two with major faults. This includes 43 samples in which typical background noises were deliberately generated. Sources of noise include the use of tools like hammering, air pressure, or electric wrenches as well as human speech, music, or ventilation units.

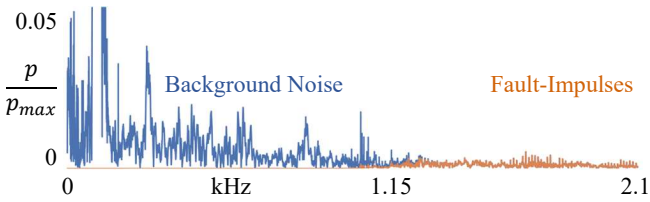

**Fig. 3:** Spectrum of a faulty motor contaminated by the noise of hammering before (blue) and after (orange) band-pass filtering.

**Data Exploration.** In line with other authors [18], we find that the frequency spectrum shows faulty impulses in a range of 850 to 5100 Hz. Disturbances induced by the inverter's switching frequency of 8 kHz, are in a higher frequency range. At the same time, acoustic disturbances have a high impact up to approximately 1150 Hz, which is demonstrated by an example in Fig. 3. Therefore, the filter frequencies are chosen as $L$ = 1150 and $H$ = 5100 Hz. However, disturbances cannot be eliminated completely while keeping fault information at the same time. Fig. 4 shows an example with and without the noise of an electric wrench. All of envelope and loudness spectra are contaminated by acoustic disturbances in the low-frequency area. For the shaft frequencies $FF_{shaft,n,k}$ four harmonics $k$ are used, which is clearly visible for the second shaft. FF and their harmonics lower than 10 Hz will not be considered. As the gear's output shaft speed is only 2.3 Hz it is not included in the features, but its gear mesh frequency and sidebands are. For the loudness spectrum, a high-pass filter with $L$ = 10 Hz and the same definition as for the band-pass filter is used to suppress the noise.

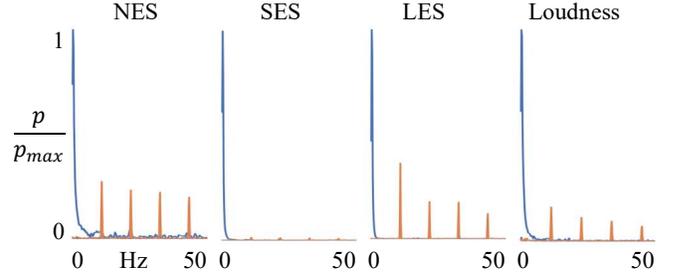

**Fig. 4:** Spectrum of a faulty motor contaminated by the noise of an electric wrench (blue) and without noise (orange).

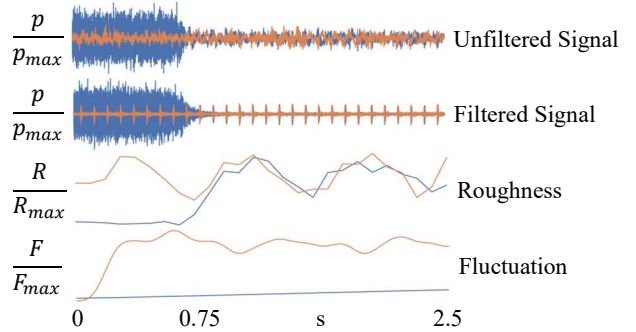

**Fig. 5:** Sample of faulty motor, where the first 0.75 s are with noise of an electric wrench (blue) and without noise (orange).

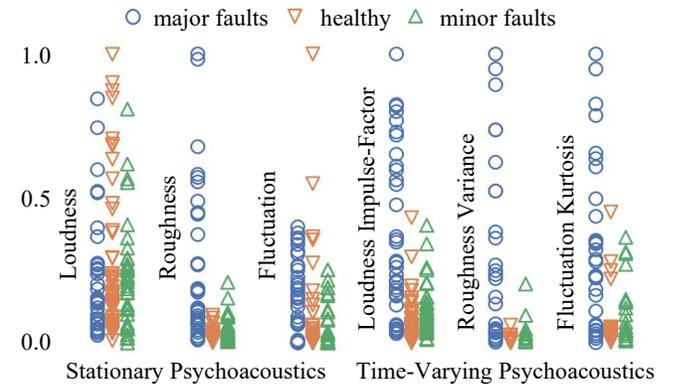

**Fig. 6:** Comparison of distributions of SPA and TVPA with scatterplots. All features are scaled relative to their maximum.

Fig. 5 shows a partly contaminated sample of a faulty motor. Fault impulses can only be observed in the filtered acoustic signal and the not contaminated area. Both SPA and TVPA are computed after applying the band-pass filter. For the roughness and loudness, the python library MOSQITO [19, 20] is used and for the fluctuation the implementation in MatLab [21]. The roughness is low for the area of noise and high otherwise, while fluctuation is low for the whole signal but high in case of no noise occurs in the entire signal. The statistical features are carefully chosen considering all types of faults and noise. The fault impulses in the loudness spectrum are described by the impulse factor, which is defined as the ratio of maximum and mean. The roughness is described by the variance, while the fluctuation is described by the kurtosis. A good performance of fault detection can be expected if a feature shows higher (or lower) values for faulty motors compared to healthy motors. However, as shown in Fig. 6 this is not the case for SPA loudness and fluctuation, but for SPA roughness. In contrast, all TVPA features show significant deviations.

## 6. RESULTS AND DISCUSSION

The performance of the different fault detection approaches is measured using a one-class-classifier (OCC), trained on healthy data. The output of the OCC is a similarity score and is used for classifying the test dataset by a threshold. [22] showed that bagging random miner (BRM) is a universal and out-of-the-box usable OCC, outperforming other well-known OCCs. The BRM [23] is trained using its default parameter. The receiver operating curve (ROC) shows the true positive rate (TPR) over the false positive rate (FPR) for all possible thresholds. Then the integral, named area under curve (AUC) describes the performance of the approaches. However, as described above, the labeling process is limited, which is demonstrated by Kendall's $W$ of 0.85. Therefore, it is suggested that an AUC of 1 may not be reachable.

**Table 1:** Performance of the fault detection approaches on LPBN. $AUC_h$ for separating the healthy motors from those with minor faults or faulty ones and $AUC_f$ for separating motors that are healthy or with minor faults from the faulty motors.

|  | NES | SES | LES | SPA | TVPA | LES+TVPA |
|---|---|---|---|---|---|---|
| $AUC_h$ | 0.68 | 0.56 | 0.84 | 0.63 | 0.76 | **0.85** |
| $AUC_f$ | 0.75 | 0.60 | 0.87 | 0.73 | 0.86 | **0.92** |

As shown in Table 1, the features of the LES outperform those of NES and SES. In related work, the SES outperforms NES for non-contaminated datasets, however, for LPBN it is the least robust approach. Therefore, we can say, the best choice of envelope type depends on data quality. Further, it is shown, that the TVPA approach outperforms the SPA significantly. The features based on TVPA outperform NES and SES, but not the LES. Combining both, LES and TVPA increase the performance. A plausible explanation is that the limitation in performance is not only caused by acoustic disturbances, but also by the different nature of the faults and labels. Then, both the envelope and psychoacoustics approaches have advantages, that the other one does not have. As seen in Fig. 4, this seems not to be the case for the loudness. Combining LES with roughness and fluctuation, but without loudness, leads to a similar performance. Therefore, it can be suggested that all relevant fault frequencies that are higher than 10 Hz are included in our approach. Looking at the ROC of the combined approach in

Fig. 7, it can be seen, that all faults can be found (low FPR) only, when many pseudo faults occur (low TPR), however, performance rises sharply then. This means some faults are included in LPBN, which have no typical fault behavior. Comparing the fault-data which has the highest OCC similarity, three out of six are contaminated by the ventilation unit. The middle part of the ROC is rising fairly, which is caused by the overlap of healthy and faulty distributions. Comparing the data with the least OCC-similarity three out of ten are contaminated by hammering and three using air-pressure.

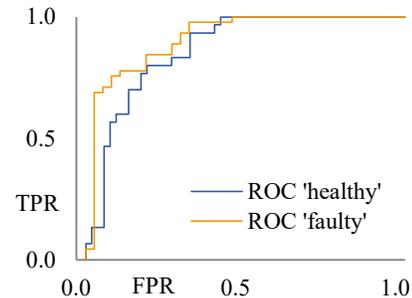

**Fig. 7:** ROC for the approach using LES and TVPA.

The problem of having faulty motors contaminated by a ventilation unit and healthy data by hammering or the use of air pressure in the same dataset is shown in Fig. 8. For the depicted example, the fault signature in roughness is much lower, but higher than those of healthy signal without noise. However, healthy data and data from minor faults could be overlapping. It can be followed that the noise of the ventilation unit potentially hides the fault impulses. Further, with the contamination of hammering or air pressure, the roughness of healthy data is rising and thus overlaps with the contaminated faulty data or with minor faults.

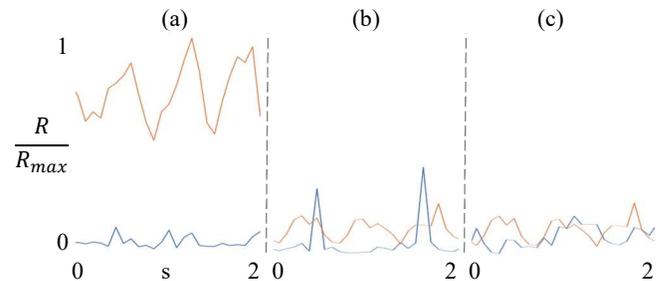

**Fig. 8:** Roughness for a healthy (blue) and a faulty (orange) motor. (a) without noise; (b) faulty: ventilation unit, healthy: hammering; and (c) faulty: ventilation unit, healthy: air pressure

## 7. CONCLUSION

We showed that acoustic gear fault detection can be applied in a real-world scenario with industrial noise, using ARAI. To face the contamination of background noise, an acoustic array and a band-pass filter were used followed by noise-robust features. It was clearly shown via data exploration and benchmark that LES and TVPA are robust approaches. LES expert features were outperforming other definitions of envelopes, which was shown for the first time for acoustic data. If expert knowledge of the gearbox construction is not available, TVPA can be used, but not SPA. The combination of LES and TVPA showed the best performance on LPBN. The most disturbing noise was the use of a hammer, air pressure, or ventilation units. Our future work will include extending LPBN with a focus on the high diversity of variants of custom-built geared motors. Further, we will explore time-frequency representation of TVPA as input for deep neural networks.